\documentclass[prl,aps,reprint,a4paper,superscriptaddress,longbibliography,floatfix]{revtex4-2}
\usepackage{graphicx}
\usepackage{epsfig}
\usepackage{amsfonts}
\usepackage{amsmath}
\usepackage{amssymb}
\usepackage{color}
\usepackage{multirow}
\usepackage[colorlinks=true,linkcolor=blue,citecolor=blue,urlcolor=blue]{hyperref}
\usepackage[utf8]{inputenc}
\usepackage[T1]{fontenc}
\usepackage{braket}

\DeclareMathOperator{\Tr}{Tr}

\newcommand{\ketbra}[2]{\ket{#1}\!\bra{#2}}

\begin{document}


\title{Complete physical characterization of QND measurements via tomography}


\author{L.~Pereira}
\email{luciano.ivan@iff.csic.es}
\affiliation{Instituto de F\'{\i}sica Fundamental IFF-CSIC, Calle Serrano 113b, Madrid 28006, Spain}
\author{J.J.~Garc\'ia-Ripoll}
\affiliation{Instituto de F\'{\i}sica Fundamental IFF-CSIC, Calle Serrano 113b, Madrid 28006, Spain}
\author{T.~Ramos}
\email{t.ramos.delrio@gmail.com}
\affiliation{Instituto de F\'{\i}sica Fundamental IFF-CSIC, Calle Serrano 113b, Madrid 28006, Spain}

\begin{abstract}
  We introduce a self-consistent tomography for arbitrary quantum non-demolition (QND) detectors. Based on this, we build a complete physical characterization of the detector, including the measurement processes and a quantification of the fidelity, ideality, and back-action of the measurement. This framework is a diagnostic tool for the dynamics of QND detectors, allowing us to identify errors, and to improve their calibration and design. We illustrate this on a realistic Jaynes-Cummings simulation of superconducting qubit readout. We characterize non-dispersive errors, quantify the back-action introduced by the readout cavity, and calibrate the optimal measurement point.
\end{abstract}


\maketitle

\begin{figure}
    \centering
    \includegraphics[width=\linewidth]{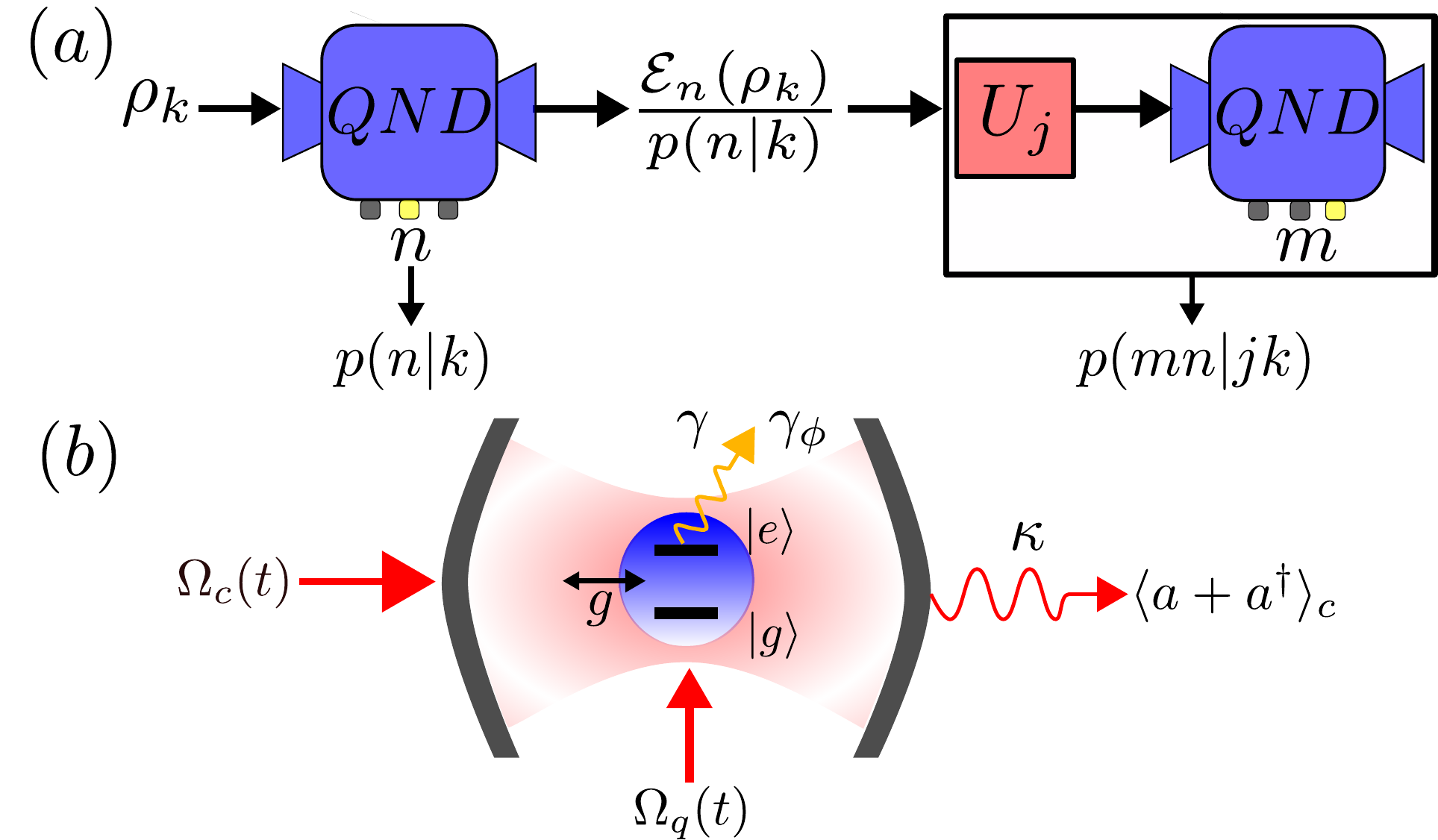}
    \caption{(a) QND detector tomography for generic measurements. A self-consistent calibration requires sampling over input states $\rho_k$ and two consecutive QND measurements, interleaved by a unitary operation $U_j$. This allows us to generate the measurement processes $\mathcal{E}_n$, for each possible outcome $n=1,\dots,N$, and to reconstruct them tomographically from the conditional probabilities $p(n|k)$ and $p(mn|jk)$. (b) Setup for tomographic characterization of QND qubit readout. It requires pulse control on qubit $\Omega_q(t)$ and cavity $\Omega_c(t)$, as well as continuous homodyne detection $\langle a+a^\dag\rangle_c$. The detector can have arbitrary qubit-cavity coupling $g$ and any imperfection such as qubit decay $\gamma$ and dephasing $\gamma_\phi$. }
    \label{fig:Fig1}
\end{figure}

\paragraph{Introduction.-} Quantum non-demolition (QND) detectors measure an observable preserving its expectation value~\cite{braginskii_quantum_1992,breuer_theory_nodate}. This property is essential in back-action-free quantum metrology~\cite{ colangelo_simultaneous_2017,moller_quantum_2017,danilishin_quantum_2012,kimble_conversion_2001,rossi_noisy_2020}, and in quantum protocols with feedback, e.g. fault tolerant computation~\cite{campbell_roads_2017,fowler_surface_2012,bermudez_fault-tolerant_2019,chamberland_building_2020,chen_exponential_2021}. QND measurements are typically implemented indirectly by monitoring a coupled subsystem, as demonstrated in various AMO~\cite{leibfried_quantum_2003,raha_optical_2020,volz_measurement_2011,gleyzes_quantum_2007,grangier_quantum_1998} and solid-state~\cite{neumann_single-shot_2010,robledo_high-fidelity_2011,nakajima_quantum_2019,xue_repetitive_2020,wallraff_approaching_2005,Walter2017} platforms. In superconducting circuits, the standard qubit measurement is a dispersive readout~\cite{blais_circuit_2020} mediated by frequency shifts in an off-resonant cavity. This is a near-QND process that approximately preserves the qubit's polarization [cf.~Fig~\ref{fig:Fig1}(b)], and enables rapid high-fidelity single-shot measurements~\cite{vijay_observation_2011,walter_rapid_2017}, protection by Purcell filters~\cite{reed_fast_2010,jeffrey_fast_2014,sete_quantum_2015,heinsoo_rapid_2018}, fast reset~\cite{mcclure_rapid_2016}, and simultaneous readout through frequency multiplexing~\cite{jerger_frequency_2012,jeffrey_fast_2014,heinsoo_rapid_2018}.

State-of-the-art QND detectors still face experimental challenges. A critical problem is the exponential accumulation of back-action errors from repeated applications of the detector, which limits the scaling of quantum technologies. In superconducting qubit readout, such errors originate in deviations from the dispersive limit in practical devices~\cite{boissonneault_dispersive_2009,Slichter2012,Sank2016}. This has motivated more complex QND measurement schemes~\cite{opremcak_measurement_2018,govia_high-fidelity_2014, siddiqi_dispersive_2006,krantz_single-shot_2016,dassonneville_fast_2020,wang_ideal_2019,didier_fast_2015,touzard_gated_2019,noh_strong_2021}, which also introduce their own sources of imperfection.

In order to make progress in the design and operation of QND measurements, we need a complete and self-consistent diagnostic tool, which helps both with the calibration of the detector and with describing its real dynamics. Many experiments have focused on optimizing simple quantities such as the readout fidelity and the QND-ness~\cite{touzard_gated_2019,dassonneville_fast_2020}. However, these fidelities do not quantify the QND nature of a measurement, but rather its projectivity and ideality as shown below. Another standard approach is detector tomography~\cite{lundeen_tomography_2009,luis_complete_1999,dariano_quantum_2004,fiurasek_maximum-likelihood_2001,Chen2019}. This method characterizes destructive measurements via positive operator-valued measurements (POVMs), but ignores the post-measurement state, and therefore a description of the measurement back-action.

In this work, we develop a complete physical characterization of QND measurements and their back-action via quantum tomography. The protocol, without pre-calibration of the QND detector, estimates both the POVM elements and the quantum process operators associated to each measurement outcome. As seen in Fig.~\ref{fig:Fig1}(a), this requires two consecutive applications of the detector, interleaved by unitary operations, and repeated over a set of input states. The information contained in the process operators can be used to identify errors, calibrate, and optimize the design of QND detectors. This can be done either directly, or through the analysis of simple metrics such as readout fidelity, QND-ness, and \textit{destructiveness}, a precise bound of the measurement back-action that we introduce below. The method can be applied to any detector, but we illustrate its power simulating realistically the calibration of superconducting qubit readout beyond the dispersive approximation. Our study shows that state-of-the-art dispersive readout is a near-ideal measurement at the optimum of QND-ness, but there are other regimes where it becomes maximally QND with minimal back-action error, as revealed by the destructiveness. Other tomographic approaches to non-destructive detectors focus on near-ideal measurements only~\cite{rudinger_characterizing_2021,blumoff_implementing_2016}, and thus do not provide a characterization of their real QND nature.

\paragraph{General description of QND measurements.-} A non-destructive quantum measurement with $N$ outcomes is represented by $N$ completely positive maps $\mathcal{E}_n$, which add up to a trace-preserving map $\mathcal{E} = \sum_n\mathcal{E}_n$. The $n$-th measurement outcome is obtained with probability $p(n) = \Tr\lbrace\mathcal{E}_n(\rho)\rbrace$, leaving the system in the post-measurement state $\rho_n = \mathcal{E}_n(\rho) / p(n)$. Each $\mathcal{E}_n$ is unambiguously represented by the Choi matrix $\Upsilon_n$, whose $d^4$ elements $\Upsilon_n^{ijkl}$ read
\begin{align}
    \Upsilon_n^{ijkl} = \bra{ij}\Upsilon_n\ket{kl} = \bra{i}\mathcal{E}_n(\ketbra{k}{l})\ket{j},
\end{align}
with $\lbrace\ket{i}\rbrace$ a basis of the measured system with dimension $d$. The Choi matrices give the post-measurement states $\mathcal{E}_n(\rho) = \sum_{ijkl}\Upsilon_n^{ijkl}\rho^{kl}\ketbra{i}{j}$~\footnote{In this work, the Choi matrix $\Upsilon_n^{ijkl}$ encodes the transformation of the density matrix~\cite{Milz2017}, while $\tilde\Upsilon_n^{ijkl}=\Upsilon_n^{ikjl}$ is a transposed version that is Hermitian and non-negative.} and also the measurement statistics $p(n)=\Tr\lbrace\Pi_n\rho\rbrace$, with the POVM elements, $\Pi_n = \sum_{ijk}\Upsilon_n^{kjki}|i\rangle \langle j|$. Conservation of probability requires the completeness relation, $\sum_n \Pi_n =\openone$, which imposes $\sum_{nk}\Upsilon_n^{kjki}=\delta_{ij}$ on the Choi components.

A QND measurement of observable $O$ ~\cite{breuer_theory_nodate,Vourdas1990,Ban1998} is one where the unconditional process $\mathcal{E}=\sum_n \mathcal{E}_n$ conserves the probability distribution $p(n)$ over repeated measurements. Equivalently, $\mathcal{E}$ preserves the average of all compatible observables $O_c$
\begin{align}
     \langle O_c \rangle=\Tr\lbrace O_c \rho \rbrace = \Tr\lbrace O_c \mathcal{E}(\rho)\rbrace, \qquad \forall [O_c,O]=0.\label{QNDcondition}
\end{align}
Ideal measurements are well known QND measurements where consecutive detections project the system onto the same eigenstate of $O$. This requires all Choi matrices to be projectors $\Upsilon_n=(\Upsilon_n)^2$, a sufficient condition to satisfy Eq.~\eqref{QNDcondition}. However, as shown below, general QND measurements are not ideal, and allow the state on consecutive detections to change, provided the averages $\braket{O_c}$ remain constant.

\paragraph{Tomographic reconstruction of measurement processes.-} We have developed a self-consistent characterization of QND measurements, based on two consecutive applications of the detector, interleaved with unitary operations from a universal set of gates $U_j$ [cf.~Fig.\ref{fig:Fig1}(a)]. The first measurement induces the processes $\mathcal{E}_n$, conditioned on the detected outcome, while the unitary $U_j$ and the second measurement are used for a process tomography of the detector itself. By repeating the protocol over a set of input states $\rho_k$, we obtain the conditional probabilities $p(n|k)=\Tr(\Pi_n\rho_k)$ and $p(mn|jk) = \Tr( \Pi_{m} U_{j}\mathcal{E}_n(\rho_k)U_{j}^\dagger)$ after the first and second measurements, respectively. From these distributions we reconstruct the POVMs $\Pi_n$ and Choi matrices $\Upsilon_n$ that best approximate the measurement, without a pre-calibration of the detector.

To recover matrices $\Pi_n$ and $\Upsilon_n$ that are meaningful and satisfy all the physical constraints of a measurement, we use maximum likelihood estimation (MLE)~\cite{ fiurasek_maximum-likelihood_2001, James2001, Shang2017} in a two-step strategy. First, we reconstruct the POVMs by minimizing the log-likelihood function
$f(\{\Pi_j\})=\sum_{n,k}\hat{p}(n|k)\log[\Tr(\Pi_n\rho_k) ]$, which compares the experimental probabilities $\hat{p}(n|k)$ to the set of feasible matrices $\lbrace \Pi_n\rbrace$ satisfying $\Pi_n\geq 0$ and $\sum_n \Pi_n =\openone$. Finally, we estimate the Choi matrices $\Upsilon_n$, minimizing a function $f_n(\Upsilon_n) = \sum_{mjk}\hat{p}(mn|jk)\log\Tr[ (U^\dagger_j\Pi_m U_j \otimes \rho_k^T)\tilde\Upsilon_n]$ which compares the experimental probabilities $\hat{p}(mn|jk)$ to a parametrization of the Choi matrix $\Upsilon_n$ satisfying $\tilde\Upsilon_n\geq 0$ ($\tilde\Upsilon_n^{ijkl}=\Upsilon_n^{ikjl}$) and the POVM constraint $\Pi_n = \sum_{ijk}\Upsilon_n^{kjki}|i\rangle \langle j|$. In total, QND detector tomography solves $N+1$ optimization problems: one for POVMs of size $d^2$, and $N$ for Chois of size $d^4$~\footnote{We solve each minimization problem using sequential least squares programming, satisfying the positivity of operators via the Cholesky decomposition, and the completeness constraints via Lagrange multipliers.}.

\paragraph{QND measurement quantifiers via tomography.-} We use the reconstructed Choi matrices $\Upsilon_n$ to quantify the performance and QND nature of a measurement. Standard benchmarks for QND detectors are readout fidelity $F=\sum_n p(n|n)/N$ and QND-ness $Q=\sum_n p(nn|n)/N$, defined as the average probability that an observable's eigenstate $\ket{n}$ remains unchanged after one or two measurements, respectively. These quantities are related to tomography via
\begin{align}
    F ={}& \frac{1}{N}\sum_n \braket{n|\Pi_n|n}=\frac{1}{N}\sum_{nj}\Upsilon_n^{njnj}, \label{eq:readout_fidelity}\\
    Q ={}& \frac{1}{N}\sum_n \braket{nn|\Upsilon_n|nn}=\frac{1}{N}\sum_n \Upsilon_n^{nnnn}.
\end{align}
The readout fidelity $F$ quantifies how close the measurement is to a \textit{projective} one $\Pi_n = (\Pi_n)^2$. QND-ness $Q$ and other similar fidelities~\cite{blumoff_implementing_2016} quantify the overlap with an \textit{ideal measurement}, satisfying $\Upsilon_n = (\Upsilon_n)^2$. Both are important measurement properties, but none of them assess the QND nature of the detector and its back-action on the observables~\eqref{QNDcondition}. For instance, a maximum value $Q=1$ characterizes an ideal measurement, but there are non-ideal measurements $Q\neq 1$ which are close to QND. Non-ideal QND measurements are useful in quantum tasks that only require evaluating observable averages such as variational algorithms or sensing \cite{suppl,Peruzzo2014,Kandala2017,Havlek2019,Thew2002,Degen2017}.

We introduce the \textit{destructiveness} $D$ as a precise quantifier of the QND nature of a detector, regardless of how ideal it is. This new quantifier bounds the change or back-action suffered by any observable compatible with the measurement of $O$, in accordance to property (\ref{QNDcondition}):
\begin{align}
D ={}& \frac{1}{2}\max_{||O_c||=1}||O_c - \mathcal{E}^\dagger(O_c)||,\qquad [O,O_c]=0.\label{eq:destructiveness}
\end{align}
Evaluating $D$ requires the tomographic reconstruction of the complete measurement process, $\mathcal{E}^\dagger(O_c)= \sum_{ijkln}\left[\Upsilon_n^{klij}\right]^\ast O_c^{kl}|i\rangle\langle j|$, and a maximization over all compatible operators of unit norm $||O_c||=1$, with $||O||=\sqrt{\Tr(O^\dagger O)}$. In practice, this maximization is done by finding the maximum eigenvalue of a positive matrix~\cite{suppl}. For instance, in the case of the qubit observable $O=\sigma_z$, the destructiveness reduces to $D = ||\sigma_z-\mathcal{E}^\dagger(\sigma_z)||/\sqrt{8}$. Since $D$ verifies the general QND condition~\eqref{QNDcondition}, it also bounds the change of the probability distribution $p(n)$ over repeated measurements.

As shown below, the three quantities $F$, $Q$, and $D$ describe the most important aspects of QND detectors, but there is further information to extract from $\Upsilon_n$, which are the most general objects.

\paragraph{Calibration of qubit readout beyond dispersive approximation.-} In standard superconducting qubit readout~\cite{blais_circuit_2020}, the qubit $\lbrace \ket{g},\ket{e} \rbrace$ couples to an off-resonant cavity mode $a$ with detuning $\Delta$ and coupling $g$ [cf.~Fig.~\ref{fig:Fig1}(b)]. For a highly anharmonic qubit, such as the flux qubit, this interaction is described by a Jaynes-Cummings (JC) Hamiltonian~\cite{Blais2004},
\begin{align}
H_\text{JC} = \frac{\Delta}{2} \sigma_z + g(\sigma_+a + a^\dagger \sigma_-) + \Omega_c(t)(a+a^\dagger),\label{JCH}
\end{align}
with Pauli operators $\sigma_z=\ketbra{e}{e}-\ketbra{g}{g}$, $\sigma_-=\sigma_+^\dag=\ketbra{g}{e}$, and a resonant drive $\Omega_c(t)$ on the cavity. In the dispersive limit $\Delta\gg g$, $H_\text{JC}$ approximates a dispersive model $H_\text{d} = \frac{1}{2}(\Delta+\chi) \sigma_z + \chi\sigma_z a^\dagger a + \Omega_c(t)(a+a^\dagger)$ that predicts a qubit-dependent displacement $\chi=g^2/\Delta$ on the cavity resonance. In theory, by continuous homodyne detection $\braket{a+a^\dag}_c$ on the cavity, we can discriminate the qubit state without destroying it. In practice, the non-dispersive corrections implicit in $H_\text{JC}$ can slightly degrade the QND nature of the measurement~\cite{boissonneault_dispersive_2009,govia_entanglement_2016}.

\begin{figure}
    \centering
    \includegraphics[width=\linewidth]{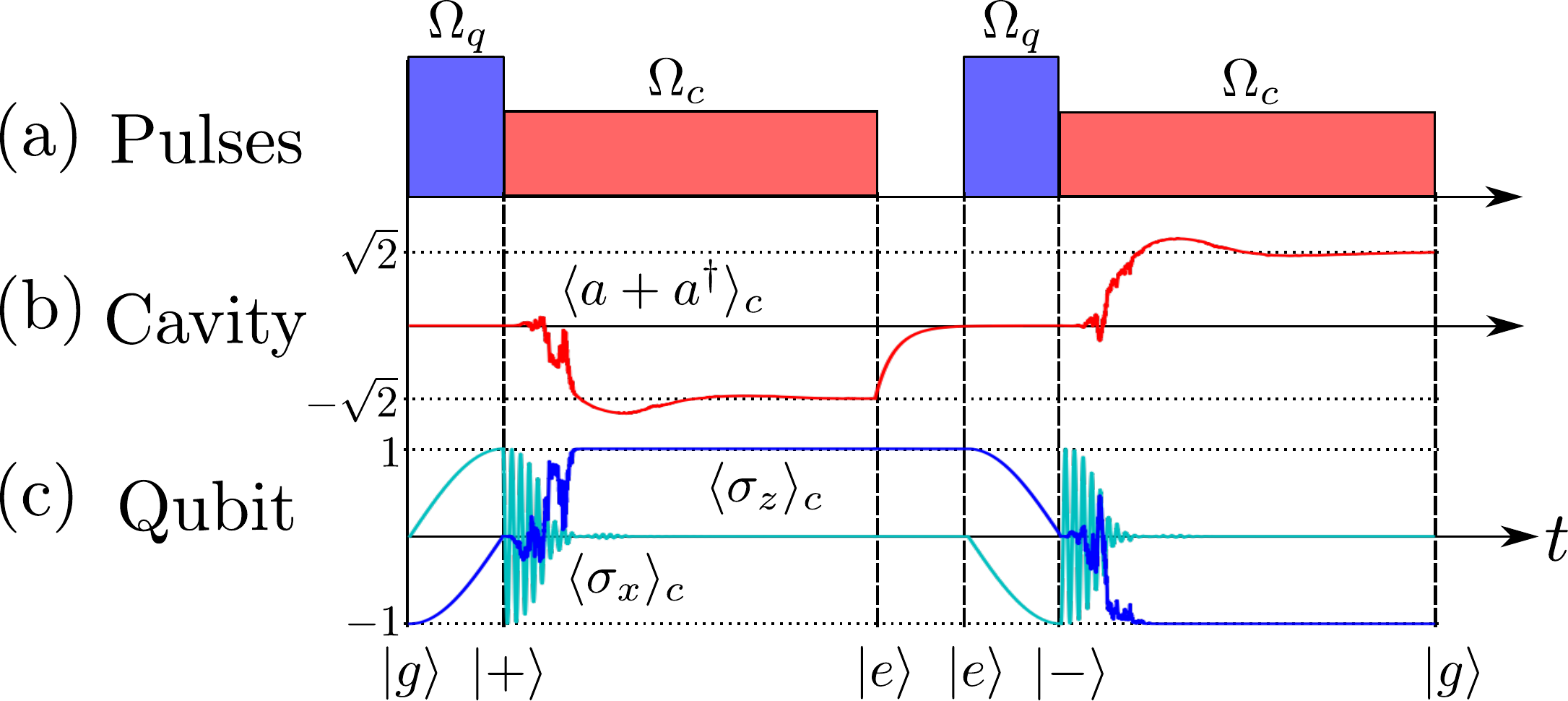}
    \caption{Simulation of QND detector tomography for dispersive qubit readout. (a) Pulse scheme on qubit (blue) and cavity (red) to implement state preparation, gates, and homodyne measurements. (b) Cavity quadrature $\langle a+a^\dag \rangle_c$ conditioned on a single trajectory. (c) Average of Pauli operators $\langle \sigma_z\rangle_c$ (blue) and $\langle\sigma_x\rangle_c$ (light blue) conditioned on the same trajectory. This realization corresponds to an initial state $|+\rangle=(|g\rangle+|e\rangle)/\sqrt{2}$ on the qubit, a first measurement with outcome $\ket{e}$, a cavity reset time, the use of gate $\exp(-i\pi\sigma_y/2)$, and a second measurement with outcome $\ket{g}$. Repeating this procedure over many trajectories with different inputs and gates allows us to reconstruct the Choi matrices $\Upsilon_g$ and $\Upsilon_e$.}
    \label{fig:pulses}
\end{figure}

To realistically quantify the performance and measurement back-action of dispersive readout, we describe the dynamics with the full $H_\text{JC}$ and a stochastic master equation (SME)~\cite{wiseman_milburn,gambetta_quantum_2008,Laflamme2017,Yang2018}. This formalism accounts for the back-action of the continuous homodyne detection onto the qubit state, as well as cavity decay $\kappa$, qubit decay $\gamma$, and qubit dephasing $\gamma_\phi$~\cite{suppl}. We simulate numerically the tomographic procedure, solving the SME over many realizations of the experiment \cite{jacobs_2010}. On each trajectory, the qubit is prepared in one of the six states $\rho_k\in\left\{ \ket{g},\ket{e},(\ket{g}\pm\ket{e})/\sqrt{2},(\ket{g}\pm i\ket{e} )/\sqrt{2} \right\}$. We perform two single-shot measurements, interleaved by a cavity reset time, and one of the three qubit gates $U_j \in \left\{ \openone, \exp(-i\pi\sigma_y/2) , \exp(-i\pi\sigma_x/2) \right\}$. In Fig.~\ref{fig:pulses} we show a representative trajectory of the protocol, where the outcome of each single-shot measurement is discriminated as $\langle \sigma_z\rangle_c =-{\rm sign}(J)$ with $J=\sqrt{\kappa}\int_0^T dt \langle a+a^\dag \rangle_c$ the homodyne current integrated over the duration $T$ of the readout pulse $\Omega_c(t)$~\cite{suppl}. For simplicity of the simulation, we neglect imperfections in the qubit state preparation and gates, performed with a local control $\Omega_q(t)$. In a real experiment, these imperfections can be self-consistently separated from intrinsic measurement errors by using standard gate set tomography~\cite{Merkel2013,Greenbaum2015,Dehollain2016}. Simulations consider state-of-the-art parameters of superconducting qubit readout~\cite{Yamamoto2014,Yan2016,walter_rapid_2017}: $g/2\pi=200 {\rm MHz}$, $\kappa=0.2g$, $\gamma=\gamma_\phi=10^{-4}g$, $T=8/\kappa\approx 32 {\rm ns}$, and $|\Omega_c| = 0.173 g$, corresponding to $\langle a^\dag a \rangle \sim 1.5$ photons on cavity for $2\chi/\kappa=1$.

\begin{figure*}[t]
    \centering
    \includegraphics[width=\linewidth]{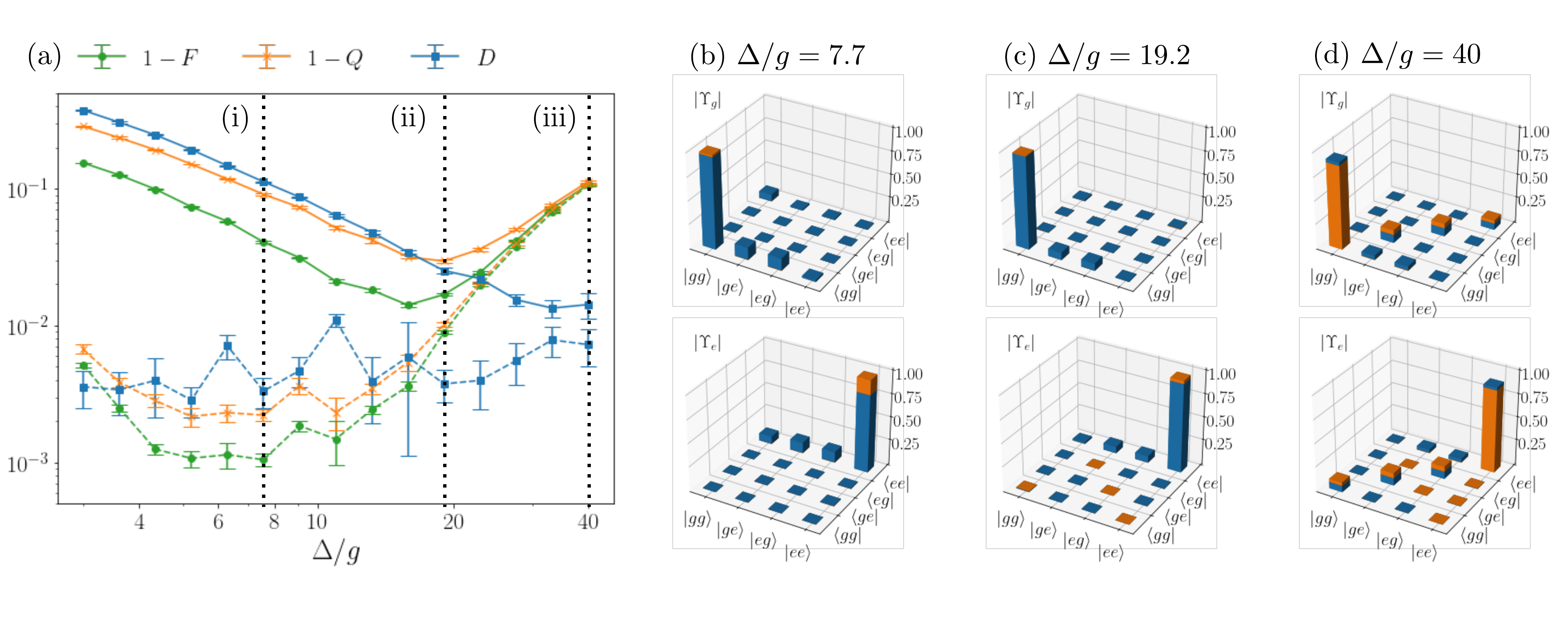}
    \vspace{-1cm}
    \caption{Measurement quantifiers and reconstructed Choi matrices for QND qubit readout. (a) Readout infidelity $1-F$ (green circles), QND-ness infidelity $1-Q$ (orange crosses), and Destructiveness $D$ (blue squares) as a function of $\Delta/g$. Solid (dashed) lines correspond to predictions for JC (dispersive) models. (b)-(d) Choi matrices $|\Upsilon_n|$ for $n=e,g$ measurement outcomes at the three representative values $\Delta/g = [7.7, 19.2, 40]$, indicated by vertical lines in (a). Blue (orange) bars correspond to predictions from JC (dispersive) models. On each column, the upper color corresponds to the part of the longest bar that does not overlap with the shortest one (and thus a single color indicates full overlap). The probabilities to reconstruct $\Upsilon_n$ are estimated from $2\times10^4$ trajectories for each initial state $\rho_k$ and gate $U_j$. Error bars correspond to one standard deviation, obtained from $10^3$ bootstrap simulations. Parameters are indicated in text.}
    \label{fig:results1}
\end{figure*}

We calibrate the measurement by tuning $\Delta/g$ and computing via tomography the quantifiers $1-F$, $1-Q$, and $D$ [cf.~Fig.~\ref{fig:results1}(a)]. We show predictions using the realistic $H_\text{JC}$ interaction (solid), as well as the dispersive model $H_\text{d}$ (dashed) to benchmark the results. We identify three qualitatively different points of operation (i)-(iii) as indicated by vertical lines in Fig.~\ref{fig:results1}(a). For each of them, we display in Figs.~\ref{fig:results1}(b)-(d) the Choi matrices $|\Upsilon_n|$ for both measurement outcomes $n=e,g$, where blue (orange) columns correspond to the JC (dispersive) predictions and the upper color corresponds to the higher values.

(i) Around $\Delta/g = 7.7$, the dispersive model reaches its optimum in fidelity and QND-ness, and the measurement is nearly ideal with projective Choi matrices $\Upsilon_n^{(d)}\approx \ketbra{nn}{nn}$ [cf.~orange columns in Fig.~\ref{fig:results1}(d)]. This occurs near $2\chi/\kappa\sim 1$ as expected by theory~\cite{blais_circuit_2020}. In contrast, the Choi matrices of the JC model show strong deviations from an ideal measurement [cf.~blue columns in Fig.~\ref{fig:results1}(b)]. The population transfer $\ket{e}\to\ket{g}$ during measurement---mainly due to cavity-induced Purcell decay~\cite{boissonneault_dispersive_2009}---reduces $\Upsilon_e^{eeee}$ and increases $\Upsilon_n^{ggee}$. In addition, the increase of $\Upsilon_n^{genn}$ and $\Upsilon_n^{egnn}$ corresponds to the growth of qubit coherences during measurement---e.g. due to cavity-mediated qubit driving \cite{govia_entanglement_2016}. These non-dispersive effects increase the infidelity and destructiveness [cf.~Fig.~\ref{fig:results1}(a)] and shift the optimal working point to larger $\Delta/g$.

(ii) Around $\Delta/g=19.2$, the non-dispersive corrections decrease and the QND-ness reaches an optimum $Q\approx 0.97$ where the actual measurement is most ideal. From $\Upsilon_n$ in Fig.~\ref{fig:results1}(c), we observe that non-dispersive effects are present but strongly suppressed. In contrast to the dispersive model, this optimum of QND-ness does not coincide with the optimum of fidelity [cf.~Fig.~\ref{fig:results1}(a)]. This is explained by noting that QND-ness is influenced by back-action $D$, while fidelity has no information about post-measurement states and cannot depend on it. In the JC model, $D$ decreases monotonically with $\Delta/g$, and thus the condition of highest $Q$ shifts towards larger $\Delta/g$ in order to minimize back-action~\footnote{Refining the discrimination criteria between measurement outcomes does not change this behavior~\cite{suppl}}. In contrast, the back-action $D$ of the dispersive model is nearly constant with $\Delta/g$, and the optima of $Q$ and $F$ coincide. 

(iii) Around $\Delta/g = 40$ the system is deep in the dispersive limit. The predictions for both models almost coincide, up to residual cavity-mediated qubit driving~\cite{govia_entanglement_2016}. Here, $1-F$ and $1-Q$ get worse, but $D$ reaches its minimum, meaning that the realistic measurement is maximally QND, but less ideal than in (ii). The loss of ideality is clearly manifested by the large diagonal terms $\Upsilon_n^{mmll}$ [cf.~Fig.~\ref{fig:results1}(d)]. This behavior is explained theoretically~\cite{suppl} by the low distinguishability between outcomes $n=g,e$ when the cavity displacement $\sim g^2/\Delta$ is too small compared to the measurement uncertainty $\sim \kappa$. The minimum value of $D$ is attributed to how decoherence $\gamma=\gamma_\phi=10^{-4}g$ breaks the QND condition~\eqref{QNDcondition} even when non-dispersive effects are suppressed~\cite{suppl}.

We see that a tomography-based calibration allows us to characterize different regimes of the detector, and to identify error sources via the process matrices $\Upsilon_n$. When simulating the measurement dynamics with $H_\text{JC}$, we account for all non-dispersive effects and the back-action appearing in realistic superconducting circuit experiments, in a unified way. We also consider imperfections due to larger intrinsic qubit decay and dephasing on the operation points (ii) and (iii). The effect of intrinsic decay $\gamma$ is qualitatively similar to Purcell decay, whereas pure dephasing $\gamma_\phi$ has a negligible effect on the Choi matrices for long readout pulses $T\gg 1/\kappa$ \cite{suppl}. Finally, our physical analysis can be used to choose an optimal working point for the detector. If one is interested in a near-ideal QND measurement, optimizing $Q$ gives a good compromise between fidelity $F$ and back-action $D$, as illustrated in (ii). This requires a larger detuning $\Delta/g$ than expected by the standard dispersive prediction \cite{blais_circuit_2020}. However, if one is interested in a maximally QND measurement with minimal back-action, $D$ is a more suitable quantity to optimize. This may require loosing ideality of the measurement as shown in (iii). 


\paragraph{Conclusions and Outlook.-} We developed a tomographic procedure to calibrate and characterize arbitrary QND detectors via a reconstruction of the measurement processes $\Upsilon_n$. We applied the method to a realistic simulation of superconducting qubit readout and identified important discrepancies between the JC and dispersive models. We expect even larger non-dispersive effects when taking into account the multi-level character of low anharmonic qubits such as transmons~\cite{Sank2016}. This is because the effective dispersive shift is reduced \cite{suppl}, and one requires lower detunings and stronger cavity drives to reach an optimal readout. While the tomography will not change, an optimized simulation of QND readout for multi-level qubits is numerically more challenging \cite{suppl} and lies outside the scope of this work.

Experimentally, the presented tomography requires only standard control and measurement tools, and therefore it can be immediately implemented to systematically analyze relevant effects on qubit readout such as the strong driving regime~\cite{boissonneault_improved_2010,Sank2016}, leakage to higher levels \cite{motzoi_simple_2009,wang_optimal_2021}, or cross-talk~\cite{hamilton_scalable_2020,bravyi_mitigating_2021}. This understanding may help improve the QND measurement performance and guide the design of alternative schemes~\cite{opremcak_measurement_2018,govia_high-fidelity_2014, siddiqi_dispersive_2006,krantz_single-shot_2016,dassonneville_fast_2020,wang_ideal_2019,didier_fast_2015,touzard_gated_2019,noh_strong_2021}. Moreover, the method can be directly applied to other platforms such as QND detectors of microwave~\cite{kono_quantum_2018,besse_single-shot_2018,lescanne_irreversible_2020,dassonneville_number-resolved_2020,essig_multiplexed_2021,curtis_single-shot_2021} or optical photons~\cite{lvovsky_continuous-variable_2009,lobino_complete_2008,zhang_mapping_2012,ramos_multiphoton_2017}. State preparation and gate errors occurring in experiments can be self-consistently included in the protocol by incorporating standard gate set tomography~\cite{Merkel2013, Greenbaum2015,Dehollain2016}. Furthermore, the reconstruction of high-dimensional Choi matrices could be done more efficiently using compressed sensing~\cite{Gross2010, Riofro2017, Ahn2019, Ahn2019_2}, matrix-product-states~\cite{Cramer2010, Lanyon2017}, or other advanced techniques~\cite{Tth2010, Goyeneche2015, Zambrano2019, Carmeli2016, Zambrano2020, 2107.05691, 2111.11071}.

From a fundamental point of view, our technique introduces an accurate procedure to quantify the back-action and real QND nature of a measurement via the destructiveness $D$. This complements the standard analysis in terms of readout fidelity and QND-ness, and allows us to identify regimes of minimal back-action regardless the ideality of the measurement. Non-ideal but highly QND measurements may be also exploited for quantum information tasks that require precise evaluations of expectation values \cite{bravyi_mitigating_2021} since the measurement outcomes can be corrected via error mitigation strategies \cite{Ban1998,botelho_error_2021,geller_conditionally_2021}.


\begin{acknowledgments}
This work has been supported by funding from Spanish project PGC2018-094792-B-I00 (MCIU/AEI/FEDER, UE) and CAM/FEDER Project No. S2018/TCS-4342 (QUITEMAD-CM). L.P. was supported by ANID-PFCHA/DOCTORADO-BECAS-CHILE/2019-772200275. T.R. further acknowledges support from the EU Horizon 2020 program under the Marie Sk\l{}odowska-Curie grant agreement No. 798397, and from the Juan de la Cierva fellowship IJC2019-040260-I.
\end{acknowledgments}


\bibliographystyle{apsrev4-2}
\bibliography{bibtex.bib}

\onecolumngrid
\newpage
{
	\center \bf \large
	Supplemental Material for: \\
	Complete physical characterization of QND measurements via tomography\vspace*{0.1cm}\\
	\vspace*{0.0cm}
}
\begin{center}
	Luciano Pereira$^{1}$, Juan Jos\'e Garc\'ia-Ripoll$^{1}$, and Tom\'as Ramos$^{1}$\\
	\vspace*{0.15cm}
	\small{\textit{$^1$ Instituto de F\'{i}sica Fundamental IFF-CSIC, Calle Serrano 113b, Madrid 28006, Spain}}
	\vspace*{0.25cm}
\end{center}

\twocolumngrid

\section*{Contents}
\begin{itemize}
	\item I.- Additional properties of Destructiveness
	\begin{itemize}
		\item I.A.- Practical calculation of Destructiveness
		\item I.B.- Destructiveness in the low back-action limit of dispersive readout
		\item I.C.- Applications of non-ideal QND measurements
	\end{itemize}
	\item II.- Stochastic master equation for qubit readout via homodyne detection on cavity
	\item III.- Choi matrices in the dispersive model
	\item IV.- Effect of qubit decoherence on Choi matrices
	\item V.- Effect of outcome discrimination criteria on Choi matrices
	\item VI.- Multi-level character of qubits in QND readout
\end{itemize}

\section{I.- Additional properties of Destructiveness}

In this section, we give a practical recipe to calculate the Destructiveness $D$ via the maximum eigenvalue of a specific positive semi-definite matrix [cf.~Sec.~I.A], a derivation of a close formula for $D$ in the case of a qubit [cf.~Sec.~I.A], and a numerical check that $D$ vanishes for qubit readout deep in the dispersive limit and for zero qubit decoherence [cf.~Sec.~I.B].

\subsection{I.A.- Practical calculation of Destructiveness}

We define the Destructiveness as a precise measure of the back-action and QND nature of a detector, given by Eq.~(5) of the main text as,
\begin{align}
	D = \frac{1}{2}\max_{||O_c||=1}||O_c - \mathcal{E}^\dagger(O_c)||,\qquad [O,O_c]=0.
\end{align}
This involves a maximization over the set of all normalized and compatible operators with $O$, and here we provide a simple recipe to calculate this quantity. Let us consider the eigenvector decomposition of $O_c=\sum_j o_j P_j$, with $P_j^2=P_j$ and $o_j\in\mathbb{R}$. Using this decomposition, the destructiveness can be rewritten as
\begin{align}
	D^2 = \frac{1}{4} \max_{ ||\boldsymbol{o}||=1 }  \sum_{jk} o_jB_{jk}o_k, \label{SM1}
\end{align}
with $B_{jk} = \Tr([P_j -\mathcal{E}^\dagger(P_j) ][P_k-\mathcal{E}^\dagger(P_k) ])$. From Eq.~(\ref{SM1}), we see that $D$ can also be calculated as the largest eigenvalue of the positive semi-definite matrix $B_{jk}$. 

In addition, it is possible to find a close expression of $D$ in the qubit case ${\cal O}=\sigma_z$, which reads
\begin{align} \label{SM1_D2}
	D=||\sigma_z-\mathcal{E}^\dagger (\sigma_z)||/\sqrt{8}.
\end{align} 
This expression is obtaining by noticing that the matrix $B_{jk}$ has at least one null eigenvalue with eigenvector $\boldsymbol{o}=(1/\sqrt{2},1/\sqrt{2})$ since the identity $O_c=I$ is compatible with any observable and this does not change with any quantum process. Using this observation in the qubit case, we find that the no-null eigenvalue of $B_{jk}$ is simply $Tr(B)$, and that its eigenvector is $\boldsymbol{o}=(1/\sqrt{2},-1/\sqrt{2})$. This eigenvector defines the compatible observable $O_c=\sigma_z/\sqrt{2}$, which solves the optimization problem \eqref{SM1}. Evaluating, we obtain Eq.~\eqref{SM1_D2}.

\begin{figure}[t!]
	\centering
	\includegraphics[width=\linewidth]{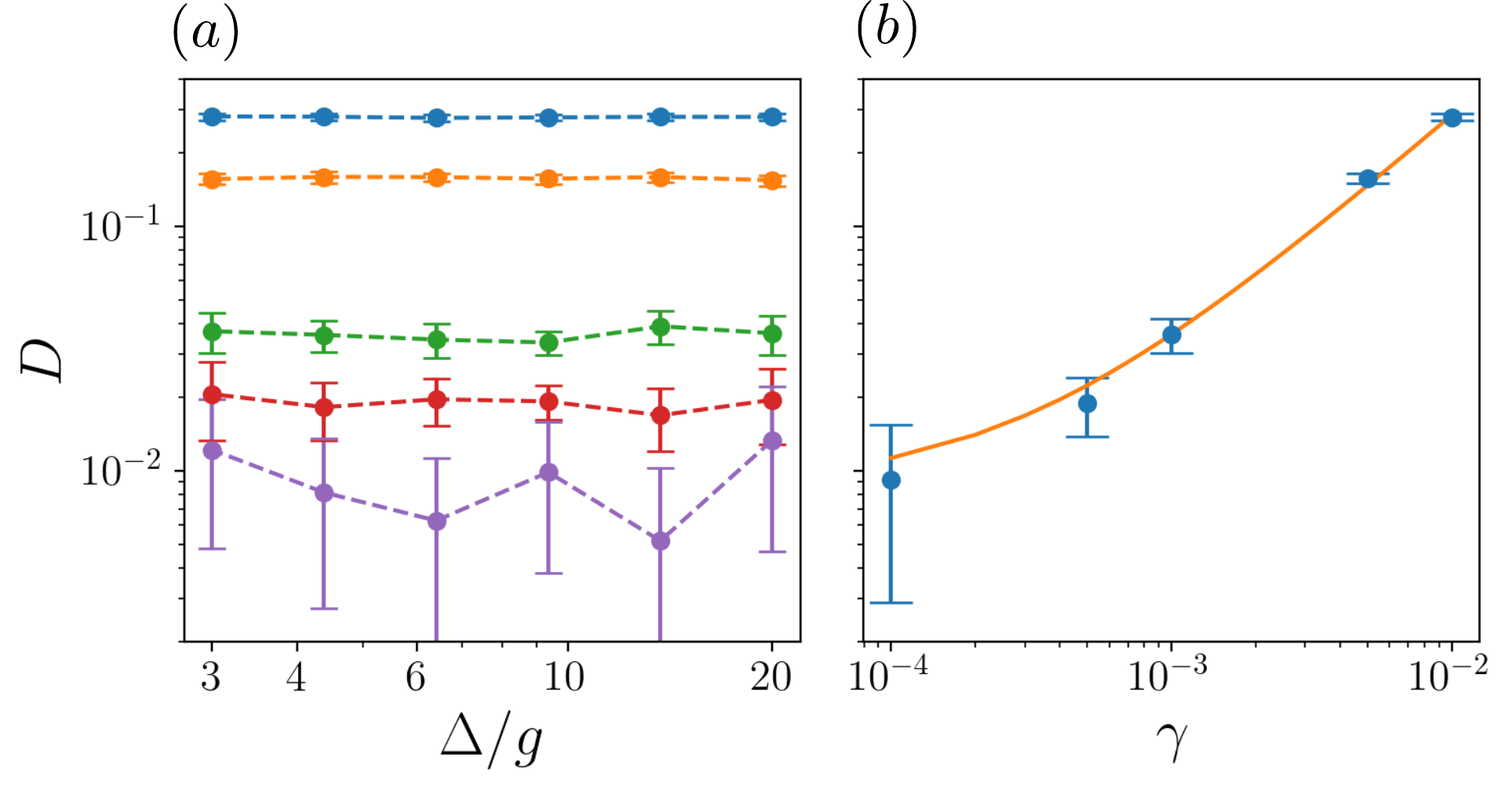}
	\caption{(a) Destructiveness $D$ obtained from QND tomography of dispersive readout, for different values of qubit decay $\gamma$. From top to bottom, $\gamma = 10^{-2}$ (blue), $5\times10^{-3}$ (orange), $10^{-3}$ (green), $5\times10^{-4}$ (red), $10^{-4}$ (purple). (b) Average Destructiveness for each value of decay. The solid orange line is a linear fit, $D=\alpha\gamma+\beta$, with $\alpha=27.7/g$ and $\beta=8.5\times 10^{-3}$. }
	
	\label{fig:fig1_SM}
\end{figure}

\subsection{I.B.- Destructiveness in the low back-action limit of dispersive readout}

In Fig.~3(a) of the main text we showed that the destructiveness $D$ does not vanish deep in the dispersive limit of qubit readout, $\Delta\gg g$, despite non-dispersive effects are strongly suppressed. Here, we show that this behavior comes from the presence of qubit decoherence in our realistic model, which breaks the exact QND condition. This occurs even in the dispersive model $H_\text{d}$, which is exempt of non-dispersive effects. 

To do so, we numerically compute $D$ by simulating the tomography via the stochastic master equation with dispersive Hamiltonian $H_\text{d}$, and for five values of the qubit decay $\gamma/g = [10^{-4},5\times 10^{-4},10^{-3},5\times 10^{-3},10^{-2}]$. For simplicity, we consider no qubit dephasing, $\gamma_\phi=0$, but the results are qualitatively similar when varying dephasing. We also consider a weak driving case, $\Omega_c=g/2$, since here we are mainly interested in the effect of qubit decay on $D$. Other parameters of the simulations are $\kappa=g/5$, and $T=10/\kappa$, and we vary the detuning $\Delta/g$. Fig.~\ref{fig:fig1_SM}(a) shows $D$ as a function of $\Delta/g$, for the five values of decay $\gamma$. These results are averaged over $10^3$ trajectories for each initial state and measurement, indicated in the main text, and the error bars correspond to $3$ standard deviations obtained from $10^3$ bootstrappings. We can see that $D$ is almost constant in $\Delta/g$ for all the decays studied. This is consistent with the dispersive model as there are no terms depending on $\Delta/g$ that may spoil the QND condition. More importantly, we see that $D$ decreases with $\gamma$, showing that $\gamma$ is responsible for at least part of the finite back-action quantified by $D$. In the dispersive model, decoherence $\gamma$ is the only quantity that breaks the QND condition, and therefore, it should be the only quantity that contributes to $D$ as well. To show this, we plot the average of $D$ as function of $\gamma$ in Fig.~\ref{fig:fig1_SM}(b), and we perform a linear fit $D(\gamma) = \alpha \gamma + \beta$, with $\alpha=27.7/g$ and $\beta=8.5\times 10^{-3}$. We observe a good agremeent between data and fit within the statistical error, which suggests that there exists a linear relationship between qubit decay and Destructiveness. Extrapolating $D$ to the decoherence-free limit, $\gamma\rightarrow 0$, we obtain that the destructiveness is lower than the statistical noise $D\le O(1/\sqrt{N_t})$, where $N_t=10^3$ is the number of trajectories used for each input state and gate in the QND detector tomography. Therefore, within the statistical precision of the tomographic reconstruction, we conclude the method predicts a vanishing destructiveness $D\rightarrow 0$, deep in the dispersive limit, and for zero decoherence.

\subsection{I.C.- Applications of non-ideal QND measurements}

The most popular QND measurement is the ideal measurement of an observable $O$, that is, projections onto the eigenspaces of $O$. They are fundamental to carry out error correction codes. It is not so well-known that non-ideal QND measurements are also useful for quantum tasks, and in particular, for algorithms that require multiple and precise evaluations of expectation values \cite{bravyi_mitigating_2021} such as variational algorithms \cite{Peruzzo2014,Kandala2017}, quantum machine learning \cite{Havlek2019}, quantum tomography \cite{James2001,Thew2002}, or quantum sensing \cite{Degen2017}. Given that the QND measurements do not change the expected value of the observable in consecutive measurements, it can be estimated several times in order to increase the sample of the observable, and then, reduce its uncertainty. However, to retrieve the expectation value of the observable of a non-ideal QND measurement it is not enough to employ the standard procedure since the POVM do not agree with the basis of eigenvectors of the observable. Instead, we have to employ a mitigation techniques \cite{Vourdas1990,Ban1998}. In the following section, we explain this procedure and show a real example of using a non-ideal QND measurement.

Let us consider a generalized measurement $\mathcal{E}_n$. Each of the measurement processes can be represented by a set of Kraus operators $K_{nm}$, such as $\mathcal{E}_n(\rho) = \sum_m K_{nm} \rho K_{nm}^\dag$. In this representation we have access to the POVM elements as $\Pi_n = \sum_{m} K_{nm}^\dag K_{nm}$ and the Choi matrices as $\Upsilon_n=\sum_m K_{nm}\otimes K_{nm}^*$. For an ideal measurement both POVM elements and Krauss operators are simply projectors $K_m=\Pi_m=\ketbra{m}{m}$, with $\ket{m}$ the basis of eigenstates of $O$. Non-ideal measurements are more general and if they are QND, they satisfy the condition,
\begin{equation}\label{eq_qnd_krauss}
	\Tr\left( O_c\sum_m K_{nm} \rho K_{nm}^\dag \right) = \Tr(O_c\rho),
\end{equation}
for any observable $O_c$ compatible with $O$. Any measurement with destructiveness $D=0$ fulfils Eq.~(\ref{eq_qnd_krauss}). In terms of the Krauss operators, a sufficient condition to satisfy the Eq.~\eqref{eq_qnd_krauss} is that $[ K_{nm}, O_c ]=0$. This kind of QND measurements also satisfy $[\Pi_n,O_c]=0$, that is, each POVM element is compatible with the observable $O_c$. This property allows us to retrieve the expected value of $O$ from $\Pi_n$ thanks to mitigation techniques even though it is not an ideal measurement of $O$ \cite{bravyi_mitigating_2021}. 

These mitigation techniques work as follows. First, if POVM elements $\Pi_n$ are compatible with $O$, we can write them as $\Pi_n = \sum_m b_{nm} \ketbra{m}{m}$. Second, let us denote $\boldsymbol{p}$ the probability distribution of an ideal measurement of $O$ and $\boldsymbol{\tilde p}$ of the non-ideal measurement $\Pi_n$, which are related by the equation $\boldsymbol{\tilde p}=B\boldsymbol{p}$, with $B$ a matrix with elements $b_{nm}$. The ideal measurement results can be obtained from the non-ideal ones by inverting the matrix,
\begin{align}
	\boldsymbol{p}=B^{-1}\boldsymbol{\tilde p}, \label{mitigationTech}  
\end{align}
allowing us to obtain the expected value of $O$. Notice that this procedure can also be applied to destructive measurements such as $[\Pi_n,O]=0$, allowing us to improve the measurement results in that case as well. However, in the case of a QND measurements the mitigation procedure also guarantees that the expectation value $\langle O \rangle$ is preserved in consecutive measurements, allowing to obtain a better precision. In this sense, the destructiveness $D$ ---introduced in this work---gives an upper bound to the change of the expected value of any observable $O_c$ compatible with $O$,
\begin{align}
	|\Tr(O_c\mathcal{E}(\rho)) - \Tr(O_c\rho)| \leq D,
\end{align}
and therefore it can be used to estimate the effectiveness of the mitigation techniques on any specific non-ideal but highly QND measurement.  

A simple example of a non-ideal QND measurement that satisfies the condition $[ K_{nm}, O_c ]=0$ is given by the following set of Krauss operators: $K_{g0}=\sqrt{1-2\epsilon}\ketbra{g}{g}$, $K_{g1}=\sqrt{\epsilon}\sigma_z$, $K_{e0}=\sqrt{1-2\epsilon}\ketbra{e}{e}$, and $K_{e1}=\sqrt{\epsilon}\sigma_z$, with $\epsilon$ a real parameter. This is a non-ideal QND measurement for any value $\epsilon < 1/2$, but in the special case of $|\epsilon|\ll 1$, it becomes a near-ideal measurement $\{ \ketbra{g}{g}, \ketbra{e}{e} \}$ affected by phase flip noise, which is a standard noise model for a realistic detector. The POVM elements of this general non-ideal measurement are $\Pi_g=\ketbra{g}{g}+\epsilon\sigma_z$ and $\Pi_e=\ketbra{e}{e}-\epsilon\sigma_z$, and its mitigation matrix reads 
\begin{align}
	B = \begin{pmatrix}
		1-\epsilon & \epsilon \\
		\epsilon & 1 - \epsilon
	\end{pmatrix}.
\end{align}
Therefore, performing the above non-ideal measurements and inverting the linear system (\ref{mitigationTech}), we can determine $\braket{O}$ as if the measurement would be ideal (for any $\epsilon<1/2$). The same mechanism applies to more complex non-ideal QND measurements.

\section{II.- Stochastic master equation for qubit readout via homodyne detection on cavity}

In this section, we explain how to model the dynamics of a dispersive qubit readout within the formalism of a stochastic Master equation (SME) \cite{wiseman_milburn, gambetta_quantum_2008, Laflamme2017}. 

A single-shot measurement is carried out by applying a resonant pulse $\Omega_c(t)$ on the cavity along a time $T$. The coherent dynamics of the measurement is given either by the Jaynes-Cummings $H=H_\text{JC}$ or the dispersive $H=H_\text{d}$ Hamiltonians, given in the main text. The cavity is continuously monitored via homodyne detection of transmitted photons, which introduces a stochastic dynamics on the qubit-cavity system. This dynamics is describe by the following SME \cite{gambetta_quantum_2008, Laflamme2017, Yang2018}
\begin{align}
	d\rho =& -i[H,\rho]dt +\kappa \mathcal{D}[a]\rho dt +\sqrt{\kappa}\mathcal{M}[a]\rho dW \nonumber \\
	& + \gamma \mathcal{D}[\sigma_-]\rho dt + \frac{\gamma_\phi}{2}\mathcal{D}[\sigma_z]\rho dt, \label{SME}
\end{align}
where the continuous homodyne measurement is described by the the Wiener process $dW$ and the super-operator $\mathcal{M}[A]\rho = (A-\langle A\rangle )\rho + \mathrm{H.c.}$. Markovian decay of the cavity $\kappa$, decay of the qubit $\gamma$, and dephasing of the qubit $\gamma_\phi$ are described by standard Lindblad super-operators, $\mathcal{D}[A]\rho = A\rho A^\dagger-\{A^\dagger A,\rho\}/2$. We numerically integrate this equation using an implicit Euler algorithm \cite{jacobs_2010}. The continuous measurement gives us access to the homodyne current over a single trajectory, 
\begin{align}
	J(t)=\sqrt{\kappa} \braket{ a+ a^\dag }_c(t)  + \xi(t),
\end{align}
where $\braket{ a + a^\dag }_c$ is the intracavity quadrature conditioned to the trajectory, and $\xi(t)=dW/dt$ is the vacuum shot-noise $\braket{\braket{\xi(t)\xi(t')}} = \delta(t-t')$. The direction of the qubit-dependent phase shift of the resonator can be determined from the homodyne current, which allows us to discriminate the state of the qubit without destroying it. To reduce the noise of the signal, we integrated $J(t)$ over the measurement time $T$, obtaining, $J=\int_0^T dt J(t)=\sqrt{\kappa}\int_0^T dt\braket{ a + a^\dag }_c$. Since the qubit-dependent cavity displacements are in opposite directions, the measurement outcome can be discriminated from the sign of the integrated signal $J$ as $\langle \sigma_z\rangle_c=-{\rm sign}(J)$.

\begin{figure}[t!]
	\centering
	\includegraphics[width=0.8\linewidth]{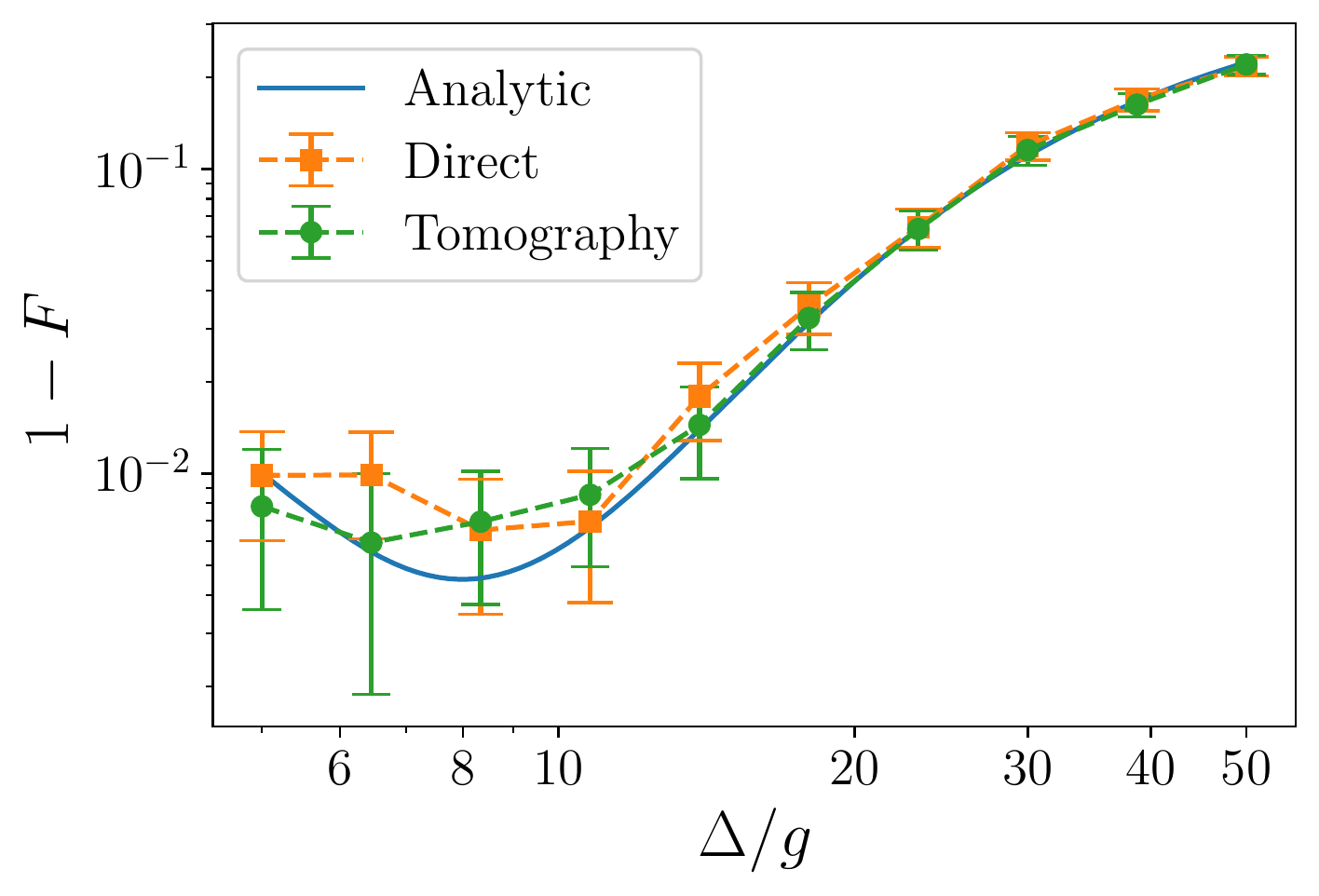}
	\caption{Comparison of three independent methods to estimate the readout infidelity $1-F$ of dispersive qubit readout. Blue corresponds to the analytical formula \eqref{analyticF}, orange to the direct definition via probablities $F = [ p(g|g)+p(e|e) ]/2$, and green to the tomographic reconstruct and the use of Eq.~(3) of the main text. The error bars correspond to 5 standard deviations obtained with $10^3$ bootstrap simulations.}
	\label{fig:fidelity_sm}
\end{figure}

In order to benchmark our code that solves Eq.~\eqref{SME}, we numerically calculate the readout fidelity $F$ via two methods. The first method consists in computing the probabilities $p(g|g)$ and $p(e|e)$ by a direct simulation of the experiment with a single qubit readout. The second method is applying our tomographic protocol involving two measurements and then obtaining $F$ via Eq.~(3) of the main text. In addition, we compare these two numerical methods with the analytical prediction in the case of the dispersive model, given in Ref.~\cite{didier_fast_2015} by
\begin{align}
	F = 1 - \frac{1}{2}{\rm erfc}({\rm SNR}/2) , \label{Festimation}
\end{align}
where ${\rm erfc}$ is the complementary error function and ${\rm SNR}$ is the signal-to-noise ratio,
\begin{align} \label{analyticF}
	{\rm SNR} = \frac{4\Omega_c \sin(\varphi)T}{\sqrt{2\kappa T}}\bigg[& 1 - \frac{4}{\kappa T }\Big( \cos(\varphi/2)^2 \nonumber \\
	& -\frac{\sin(\chi T+\varphi)}{\sin(\varphi) }e^{-\kappa T/2} \Big) \bigg]. 
\end{align}
Here, $\varphi=2\arctan(2\chi/\kappa)$. For the simulations, we set the parameters on $g$ units as $\kappa/g=1/5$, $\gamma=\gamma_\varphi =0$, $\Omega_c/g=1/10$, and $T=10/\kappa$. The average number of photons in the resonator is $\braket{a^\dag a}=0.5$. We consider here a weak driving case since it requires a low number of trajectories to achieve a good precision in the numerical calculations. For each initial state and gate needed for the tomography (indicated in the main text), we perform $10^3$ trajectories. The error bars are 5 times the standard deviation obtained with $10^3$ bootstrap simulations. The results are shown in Fig.~\ref{fig:fidelity_sm}. We see that the three independent estimations of $F$ ---direct, tomography, and analytical--- agree well within an error of 5 standard deviations.

\section{III.- Choi matrices in the dispersive model}

In this section, we study the general structure of Choi matrices $\Upsilon_n$ predicted by the dispersive model. With this we show that the diagonal form of $\Upsilon_n$ in Fig.~3(d) of the main text is characteristic from a low measurement indistinguishably occurring deep in the dispersive limit. 

To do so, let us consider that the qubit is initially in the state $\ket{\Psi} = \psi_e\ket{e}+\psi_g\ket{g}$ and the cavity is empty $|0\rangle_r$. Modeling the system with the dispersive Hamiltonian $H_\text{d}$, the coherent evolution of the joint system $\ket{\Psi} = \ket{\Psi}_q|0\rangle_r$ after a time $t$ is given by \cite{blais_circuit_2020}
\begin{align} \label{qubit_cavity_state}
	\ket{\Psi(t)} = \psi_g\ket{g}\ket{\alpha_g(t)} + e^{-i\Delta t}\psi_e\ket{e}\ket{\alpha_e(t)}.
\end{align}
Here, $|\alpha_j(t)\rangle$ is the resonator state whose phase is shifted conditioned on the qubit state $j\in\{g,e\}$. Therefore, we can infer the qubit state by measuring the cavity field. This evolution is described by the unitary operation,
\begin{align}
	U(t) = \ketbra{g}{g}\otimes D(\alpha_g(t)) + e^{-i\Delta t}\ketbra{e}{e}\otimes D(\alpha_e(t)),
\end{align}
where $D(\alpha)$ is the cavity displacement operator. Reading the qubit state requires continuous homodyne detection during a time $T$ of the cavity quadrature operator $\tilde{Q} = \sqrt{\kappa}\left(  a + a^\dagger \right)$. The outcome of the qubit measurement is discriminated form the integrated homodyne current, $J = \frac{1}{T}\int_0^Tdt\langle\Psi(t)| \tilde{Q} \ket{\Psi(t)}$. This can be obtained as the expected value of $\tilde{Q}$ in the time average density matrix, $\mathcal{E}_{qr}(\ketbra{\Psi}{\Psi}) = \frac{1}{T} \int_0^T dt\ketbra{\Psi(t)}{\Psi(t)}$. In addition, the reduced evolution for the qubit can be obtained by tracing over the cavity $\mathcal{E}_q(\ketbra{\Psi}{\Psi}) = \Tr_r[ \mathcal{E}_{qr}(\ketbra{\Psi}{\Psi}) ]$. Considering the eigenvector basis of $\tilde{Q}$, $\{|q\rangle\}$, we have that the Kraus representation of the qubit dynamics is
\begin{align}
	\mathcal{E}_q(\ketbra{\Psi}{\Psi}) = \int_0^T dt \int_{-\infty}^\infty dq  K_q(t)\ketbra{\Psi}{\Psi} K_q(t)^\dagger,
\end{align}
with Kraus operators $K_q(t) = {}_r\langle q|U(t)|0\rangle_r/\sqrt{T}$ \cite{breuer_theory_nodate}. Considering the wave function of the cavity states $\phi_j(q,t)=\langle q | \alpha_j(t)\rangle$, the Kraus operators can be expressed as
\begin{align}
	K_q(t) = \frac{1}{\sqrt{T}}( \phi_g(q,t) \ketbra{g}{g} + \phi_e(q,t)e^{-i\Delta t}\ketbra{e}{e} ).
\end{align}

We can divide the trajectories in order to determine the outcome of the measurement. If the homodyne current is larger than a predefined value $\delta$, we assign the result $e$, and otherwise we assign $g$. For instance, in the special case of a dispersive model without decoherence both states displace the cavity by the same amount, but in opposite directions, having $\delta=0$. Using this separation in the general case, the components of the Choi matrices are given by
\begin{align}
	\Upsilon_g^{ijkl} =&  \int_0^Tdt\int_{-\infty}^\delta dq \bra{i}K_q(t)\ketbra{k}{l}K_q(t)^\dagger\ket{j}, \\
	\Upsilon_e^{ijkl} =&  \int_0^Tdt\int_\delta^{\infty} dq \bra{i}K_q(t)\ketbra{k}{l}K_q(t)^\dagger\ket{j}. 
\end{align}
We can further simplify the Choi matrices as, 
\begin{align}\label{eq:choi_g}
	\Upsilon_g = & \epsilon_g\ketbra{gg}{gg} + (1-\epsilon_e)\ketbra{ee}{ee}\nonumber\\
	& + \zeta_g \ketbra{ge}{ge} + \zeta_g^*\ketbra{eg}{eg},
\end{align}
\begin{align}\label{eq:choi_e}
	\Upsilon_e = & (1-\epsilon_g)\ketbra{gg}{gg} + \epsilon_e\ketbra{ee}{ee}\nonumber\\
	& + \zeta_e^* \ketbra{ge}{ge} + \zeta_e\ketbra{eg}{ge},
\end{align}
where we have defined the four auxiliary quantities $\epsilon_g =\frac{1}{T} \int_0^Tdt\int_{-\infty}^\delta dq|\phi_e(q,t)|^2$, $\epsilon_e = \frac{1}{T}\int_0^Tdt\int_\delta^{\infty}dq|\phi_g(q,t)|^2$, $\zeta_g = \frac{1}{T}\int_0^Tdt\int_{-\infty}^\delta dq e^{i\Delta t}\phi_g(q,t)\phi_e(q,t)^*$, and $\zeta_e = \frac{1}{T}\int_0^Tdt\int_\delta^{\infty}dq e^{-i\Delta t}\phi_e(q,t)\phi_g(q,t)^*$.

The general form of $\Upsilon_n$ in Eqs.~\eqref{eq:choi_g}-\eqref{eq:choi_e} deviate from projectors and thus we conclude that the measurement is not ideal in general. In the case without decoherence, the wave functions correspond to coherent states. In this case, the outcomes of the measurement cannot be perfectly discriminated since the coherent states are not orthogonal. Therefore, we have that $\epsilon<1$ and $|\zeta_j|>0$, and that the dispersive readout cannot be a perfectly ideal measurement. When the cavity displacement $\sim g^2/\Delta$ is too small compared to the measurement uncertainty $\sim \kappa$, the overlap between the wave functions increase, implying that $\epsilon_j$ decrease and $|\zeta_j|$ increase. We can see this effect in the Choi matrices of Fig.~3(d) of the main text, which exhibits large diagonal terms $\Upsilon_n^{gege}$ and $\Upsilon_n^{egeg}$. 

\section{IV.- Effect of qubit decoherence on Choi matrices}

In this section, we simulate the tomographic characterization of dispersive readout as a function of qubit decay $\gamma$ and dephasing $\gamma_\phi$ to study its effect on the Choi matrix components $\Upsilon_g$ and $\Upsilon_e$.

\begin{figure}[t]
	\centering
	\includegraphics[width=0.8\linewidth]{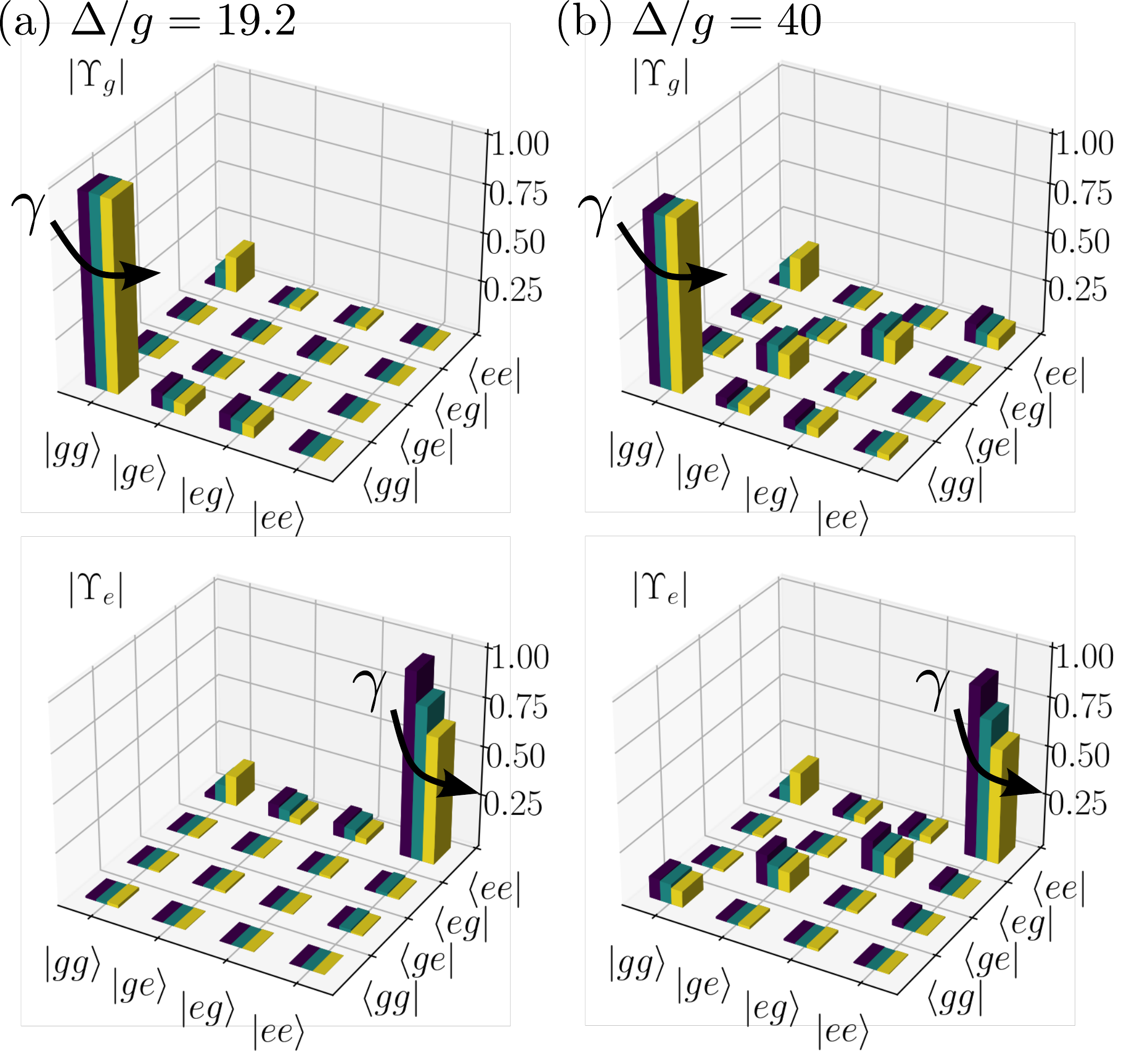}
	\caption{Choi matrices of dispersive readout with JC model in presence of decay $\gamma$ for (a) $\Delta/g=19.2$ and (b) $\Delta/g=40$. 
		Columns of different colors represent different values of decay, which increase in the direction indicated by the arrow as $\gamma/g=10^{-4}$ (blue), $5\times10^{-3}$ (green), $10^{-2}$(yellow).}
	\label{fig:decoherence1}
\end{figure}

\begin{figure}[t]
	\centering
	\includegraphics[width=0.8\linewidth]{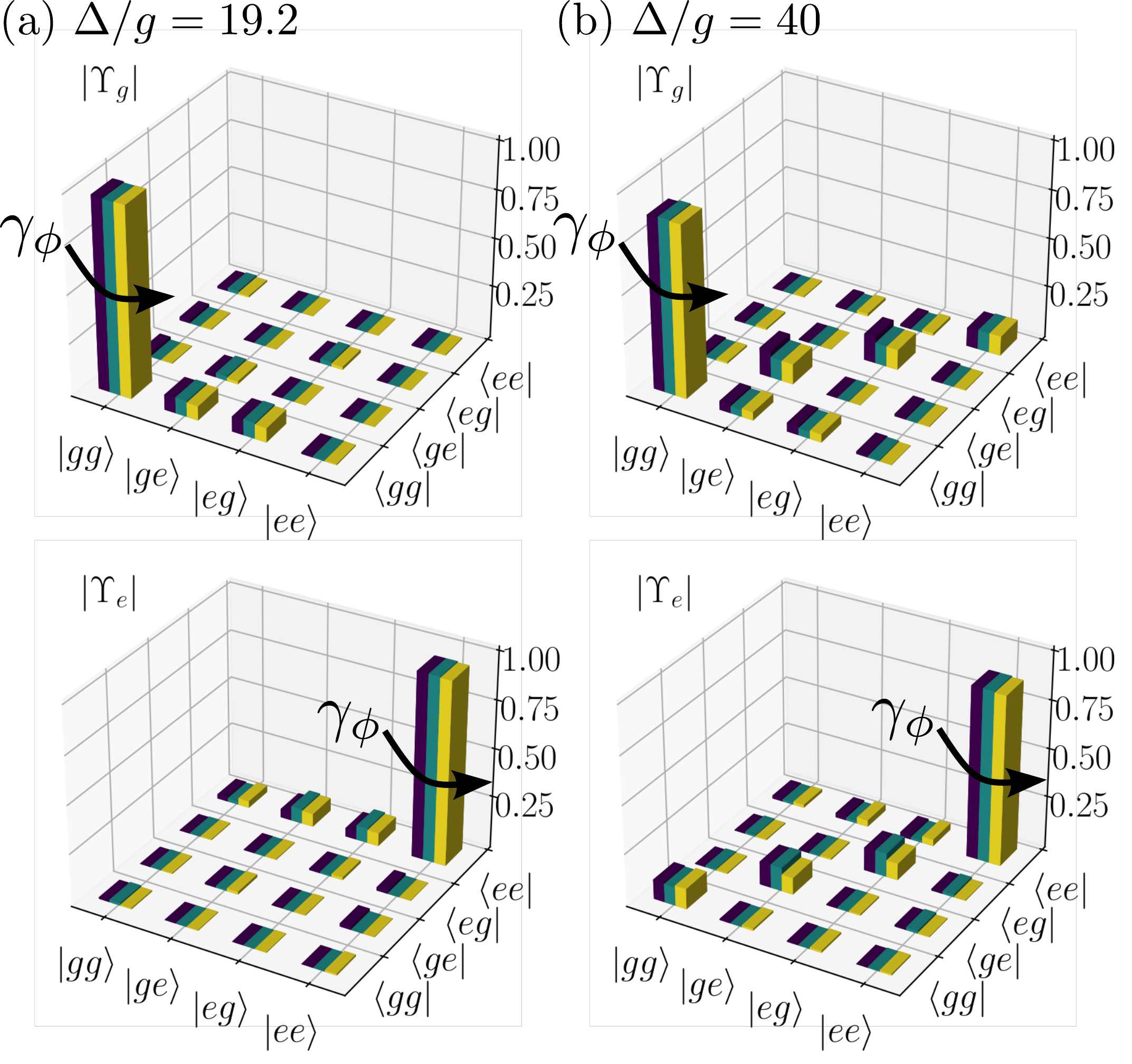}
	\caption{Choi matrices of dispersive readout with JC model in presence of dephasing $\gamma_\phi$ for (a) $\Delta/g=19.2$ and (b) $\Delta/g=40$. 
		Columns of different colors represent different values of dephasing, which increase in the direction indicated by the arrow as $\gamma_\phi/g=10^{-4}$ (blue), $5\times10^{-3}$ (green), $10^{-2}$(yellow).}
	\label{fig:decoherence2}
\end{figure}

The simulations of QND detector tomography are done using the JC Hamiltonian, and with $N_t=2\times 10^3$ trajectories for each initial state and gate (indicated in the main text). Parameters are $\kappa/g=1/5$, $T=6/\kappa$, and $\Omega_c/g=\sqrt{3}/10$.
For $\Delta/g$, we consider the operating points (ii) and (iii) of Fig.~3(a) of the main text, that is a near-ideal measurement $\Delta/g = 19.2$, and the deep dispersive limit $\Delta/g=40$, respectively. For both cases, we compute the Choi matrices $|\Upsilon_n|$ for three values of decay $\gamma/g$ as shown by columns of different colors in Fig.~\ref{fig:decoherence1}. Similarly, we consider three values of dephasing $\gamma_\phi/g$, and the corresponding $|\Upsilon_n|$ are shown in Fig.~\ref{fig:decoherence2}. Both operating points (ii) and (iii) show similar behavior when increasing qubit and dephasing, indicating that decoherence manifests in $\Upsilon_n$ independently of $\Delta/g$. From Fig.~\ref{fig:decoherence1} we see that the main effect of decay $\gamma$ is to increase the Choi components $\Upsilon_n^{ggee}$ and decrease $\Upsilon_n^{eeee}$. This behavior is consistent with the effect of cavity-induced Purcell decay discussed in the main text. From Fig.~\ref{fig:decoherence2} we see that the dephasing has marginal effect on the Choi matrix because all the deviations in its components are on the order of magnitude or lower than the statistical error $O(1/\sqrt{N_t})$. This is reasonable since dephasing cannot appreciably affect the dynamics of the qubit after it has been projected to $\ket{g}$ or $\ket{e}$ by the measurement. Dephasing may be relevant for short measurement times $T$, when the projection has not yet fully happened, but in all our simulations we have considered $T\gtrsim 6/\kappa$. 

\section{V.- Effect of outcome discrimination criteria on measurement quantifiers}

The simulation of QND readout carried out in the main text uses a simple discrimination strategy based on the sign of the integrated homodyne current (cf.~Sec.~II of SM). In this section, we consider two ways of refining the discrimination method between outcomes and we calculate its effect on the quantifiers $F$, $Q$, and $D$ obtained from the QND detector tomography. A first refining comes from noticing that the homodyne current at short times is not enough to discriminate the outcome of the measurement, and therefore it is better to employ an integrated homodyne current that ponderates more the values at longer times instead of using a flat integral such as in Sec.~II of the SM. This can be done by introducing a weighting function $w(t)$ in the determination of the integrated homodyne current \cite{walter_rapid_2017} as
\begin{align}
	J = \int_0^T dt w(t) J(t),
\end{align}
with $J(t)=\langle a+a^\dagger \rangle_c$ the time-dependent homodyne current of a single trajectory and $T$ the integration time. A second simple refinement comes from noticing that when non-dispersive effects are included, the displacement of the cavity by the ground and the excited states are not symmetric. Specifically, the displacement of the excited state is lower than the ground state. Because of this, one obtains a better discrimination of outcomes when placing the discrimination threshold on the integrated homodyne at a finite value $\delta$ instead of simply at zero, as done in Sec.~II of the SM and in the main text. Both the weighting function $w(t)$ and threshold $\delta$ can be calibrated from homodyne currents $J_g(t)$ and $J_e(t)$ obtained from solving the exact master equation when the qubit is initially in $\ket{g}$ or $\ket{e}$, respectively. In particular, they are given by $w(t) = |J_g(t) - J_e(t)|$ and $\delta = (J_g(T)+J_e(T))/2$. In a real experiment, it is not necessary to perform these simulations to calibrate $w(t)$ and $\delta$ because they can also be determined from a set of experimental trajectories \cite{walter_rapid_2017}.

\begin{figure}
	\centering
	\includegraphics[width=\linewidth]{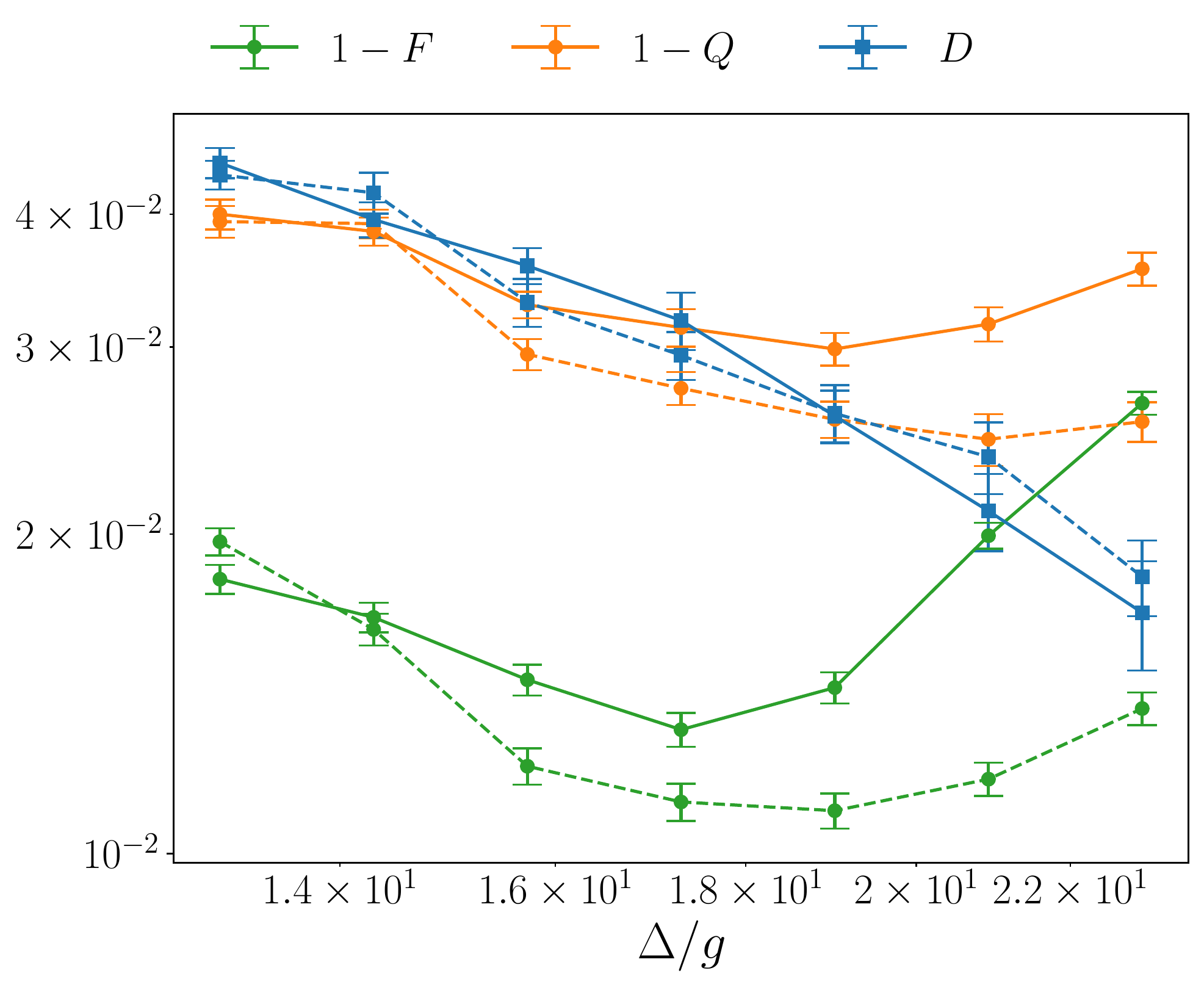}
	\caption{Readout infidelity $1-F$ (green circles), QND-ness infidelity $1-Q$ (orange crosses), and Destructiveness $D$ (blue squares) as a function of $\Delta/g$ for QND qubit readout in the case of two different discrimination criteria: Solid lines correspond to predictions with the simple discrimination method used in the main text, while dashed lines correspond to the optimized discrimination method with nonzero threshold and weighting function. In both cases, simulations are done with JC model and for the same parameters as in the main text. Probabilities were estimated from $10^4$ trajectories for each initial state and gate necessary for the tomography. Error bars correspond to one standard deviation, obtained from $10^3$ bootstrap simulations.}
	\label{fig:optimized measurement}
\end{figure}

In Fig.~\ref{fig:optimized measurement} we compare the three measurement quantifiers $F$, $Q$, and $D$ obtained from the tomography using either the simple discrimination of the main text (solid) or the optimized discrimination explained above (dashed). For the simulations we consider the full JC model with the same parameters of the main text. Our main conclusion is that refining the discrimination criteria improves the overall quality of the measurement since the readout and QND-ness infidelities reduce, but the specific features of the measurement remain unchanged. In particular, for both discrimination strategies, the minimum of $1-Q$ occurs at a larger detuning $\Delta/g$ than the minimum of $1-F$. The effect of improving the discrimination is that both minima move towards larger detuning $\Delta/g$ but they keep their distance roughly the same on order $\Delta/g\sim 2$ (the minimum of $1-F$ moves from $\Delta/g\sim 17$ to $\Delta/g\sim 19$, while the minimum of $1-Q$ moves from $\Delta/g\sim 19$ to $\Delta/g\sim 21$). As discussed in the main text, these difference of the minima of $1-F$ and $1-Q$ has to do with the back-action introduced by the cavity in the JC model, which reduces at larger $\Delta/g$ as quantified by $D$. Finally, it is also interesting to notice that the destructiveness $D$ is not appreciably changed when improving the discrimination criteria, which is consistent for a quantifier of the physical back-action of a measurement since this cannot depend on the post-processing strategy used to interpret the data.

\section*{VI.- Multi-level character of qubits in QND readout}

Superconducting qubits are more precisely modeled as multi-level systems with Hamiltonian $H_q=\sum_k \omega_k \ket{k}\bra{k}$, where $\omega_k$ is the resonance frequency of each of their eigenstates $\ket{k}$ ($k=0,1,2,\dots$) \cite{blais_circuit_2020,Yan2016,Sank2016}. The relevance of the multi-level character of qubits depends on the size of their anharmonicity $\alpha_q=(\omega_2-\omega_1)-(\omega_1-\omega_0)$ compared to the other external control fields applied on the qubit as we see below.

To include the multi-level character of qubits in the simulation of QND readout, one must generalize the JC Hamiltonian of the main text to a multi-level JC Hamiltonian \cite{Sank2016},
\begin{align}
	H_\text{JC}^{\rm{(ML)}} ={}& \sum_k \omega_k \ket{k}\bra{k} + \omega_c a^\dag a + \Omega_c(ae^{i\omega_d t}+a^\dagger e^{-i\omega_d t})\nonumber\\ 
	{}&+ \sum_{k\neq k'}g_{kk'}\ket{k'}\bra{k}(a+a^\dag).\label{HJCML}
\end{align}
Here, $\omega_c$ is the cavity frequency, $\omega_d$ the driving frequency, $\Omega_c(t)$ the cavity driving strength, and $g_{kk'}$ the qubit-cavity couplings for the qubit transition $\ket{k}\rightarrow\ket{k}'$ (satisfying $g_{kk'}=g_{k'k}$). Notice that Eq.~(\ref{HJCML}) describes qubit-cavity interaction processes outside the rotating wave approximation (RWA) which is reported to be relevant when the cavity is highly populated $\braket{a^\dag a}\gg n_c=(g/2\Delta)^2$ \cite{Sank2016}. The dissipative part of the dynamics is governed by the same stochastic master equation presented in Sec.~II of this SM, replacing the Lindblad terms of $\sigma_-$ and $\sigma_z$ by the various transition operators $\ket{k'}\bra{k}$ ($k\neq k'$) and projectors $\ket{k}\bra{k}$, respectively. 

The Hamiltonian (\ref{HJCML}) describes the complete coherent dynamics of the standard $\ket{g}=\ket{0}$ and $\ket{e}=\ket{1}$ states of the qubit, but also includes all other higher excited states $\ket{k}$ with $k\geq 2$. However, if one is only interested in a QND measurement of the two-level manifold $\lbrace \ket{g},\ket{e} \rbrace$ and in the moderate driving regime ($\braket{a^\dag a} \ll n_c$), the most relevant deviation from the naive two-level approximation comes from the second excited level $\ket{f}=\ket{2}$. In this case, we can neglect the non-RWA terms and the JC multi-level Hamiltonian (\ref{HJCML}) can be approximated by a qutrit Hamiltonian, which in the rotating frame with the driving frequency $\omega_d$ can be written as:
\begin{align}
	H_\text{JC}^{\rm{(3)}} ={}& \frac{\Delta_q}{2} \sigma_z + \frac{3}{2}\left(\Delta_q+\alpha_q\right)|f\rangle\langle f|+ \Delta_c a^\dag a\nonumber\\ 
	{}& +g(\sigma^+ a + a^\dagger \sigma^-) + \sqrt{2}g\left(\ket{f}\bra{e}a+a^\dag \ket{e}\bra{f}\right) \nonumber\\
	{}&+ \Omega_c(t)(a+a^\dagger).\label{DHT3}
\end{align}
Here, $\sigma_-=\ket{g}\bra{e}$ and $\sigma_z=\ket{e}\bra{e}-\ket{g}\bra{g}$ are the Pauli operators, $\Delta_q=(\omega_{1}-\omega_0)-\omega_d$ is qubit-drive detuning, $\Delta_c=\omega_c-\omega_d$ the cavity-drive detuning, $\alpha_q=(\omega_2-\omega_1)-(\omega_1-\omega_0)$ the qubit anharmonicity, and we simplified the notation for the couplings as $g_{01}=g_{10}=g$ and $g_{12}=g_{21}=\sqrt{2}g$. To gain insight on the main effects of the extra $\ket{f}$ level on the QND readout scheme, we apply standard perturbation theory on Eq.~(\ref{DHT3}) \cite{blais_circuit_2020} and project onto the $\lbrace \ket{g},\ket{e} \rbrace$ manifold, obtaining a modified dispersive Hamiltonian $H_\text{D}'$ for an approximated two-level system:
\begin{align}
	H_\text{D}' ={}& \frac{1}{2}(\Delta+\Delta_c+\chi') \sigma_z + \left(\Delta_c-\frac{g^2}{\Delta+\alpha_q}\right)a^\dag a \nonumber\\ 
	{}&+\chi' \sigma_z a^\dagger a + \Omega_c(t)(a+a^\dagger).\label{DHT}
\end{align}
This two-level model is valid in the dispersive regime $g\ll |\Delta|, |\Delta+\alpha_q|$, with $\Delta=(\omega_1-\omega_0)-\omega_c$ the qubit-cavity detuning. Setting $\Delta_c=g^2/(\Delta+\alpha_q)$ to compensate the cavity shift, Eq.~(\ref{DHT}) has exactly the same form as the standard dispersive Hamiltonian used in the main text, except that here the dispersive shift $\chi'$ is modified by the finite anharmonicity $\alpha_q$ of the qubit as \cite{walter_rapid_2017}
\begin{align}
	\chi'=g^2\left(\frac{1}{\Delta}-\frac{1}{\Delta+\alpha_q}\right)=\frac{g^2}{\Delta}\frac{\alpha_q}{\Delta+\alpha_q}.\label{modifiedShift}
\end{align}
Therefore, for a supeconducting qubit with anharmonicity much larger than the detuning $|\alpha_q|\gg|\Delta|$, we recover the dispersive model of the main text with $\chi=g^2/\Delta$ as the $\ket{f}\leftrightarrow\ket{e}$ transition is much more off-resonant than $\ket{e}\leftrightarrow\ket{g}$. However, for low anharmonic qubits (such as transmons \cite{walter_rapid_2017,Sank2016}), one typically has $|\alpha_q|\lesssim|\Delta|$ (with $\alpha_q<0$) and therefore the main effect of the $\ket{f}$ state is to reduce the magnitude of the dispersive shift by a factor $\alpha_q/(\alpha_q+\Delta)$ compared to the naive two-level model. Importantly, one can reduce $\Delta$ to obtain the same effective shifts as predicted by the naive two-level model. For instance, for $\alpha_q<0$ and $\Delta<0$, if we choose $\Delta'=|\alpha_q|/2+ \sqrt{|\alpha_q|^2/4+|\Delta||\alpha_q|}$ we get $|\chi'(\Delta')|=|\chi(\Delta)|$ (the sign of the shift does not affect the measurement performance). This means that results for the dispersive model of Fig.~3 of the main text (obtained in the range $|\Delta|/g\in [3,40]$) can be re-scaled to account for the most important multi-level effects of the qubit, but at detunings in the shorter range $\Delta'/g\in [3,9]$ \footnote{Notice that the calculations in the main text using the two-level model do not depend on the sign of $\Delta$, but the inclusion of the $|f\rangle$ state breaks this symmetry}. The above is valid deep in the dispersive regime ($g\ll |\Delta|, |\Delta+\alpha_q|$) and at moderate drive power ($\braket{a^\dag a}\ll n_c$), but studying non-dispersive effects in the multi-level model is much more complex than for a two-level system \cite{Sank2016}.

In practice, when performing QND readout of low anharmonic qubits, we expect that deviations from the dispersive model $H_D'$ should become more important than predicted by the two-level model in Fig.~3 of the main text. This is because the multi-level character of the qubit reduces the effective dispersive shift $\chi'$ as indicated in Eq.~(\ref{modifiedShift}), and therefore one needs to perform the measurement at lower detunings $\Delta$ and larger cavity drives $\Omega_c$ to obtain a similar resolution for discriminating the outcomes $n=g,e$. For instance, for a transmon qubit with a typical anharmonicity of $\alpha_q/2\pi\sim 300$ MHz, we estimate using Eq.~(\ref{Festimation}) that reaching an optimal measurement performance such as for point (ii) in Fig.~3, we require to reduce the detuning $\Delta/g$ by a factor of $3$, as well as to increase the product $\Omega_c T$ by a factor of $2.5$ (with $T$ the measurement time). These two changes in parameters make the dispersive approximation ($g\ll |\Delta|, |\Delta+\alpha_q|, \braket{a^\dag a}\ll n_c$) less valid and therefore the non-dispersive effects should be more pronounced. 

Exactly quantifying these non-dispersive effects with a full tomography of multi-qubit QND readout using Hamiltonians (\ref{HJCML}) or (\ref{DHT3}) is numerically very challenging. This is not only because the qubit has a larger dimension, but also because one needs to account for many more photons in the dynamics of the cavity due to the increase of $\Omega_c T$ to compensate the reduction of $\chi'$. We estimate that increasing $\Omega_c T$ by a factor 2.5 with respect to what was used in Fig.~3, implies duplicating the integration time from $T=8/\kappa$ to $T=16/\kappa$, as well as increasing the steady-state cavity population $\braket{a^\dag a}$ from 1.5 to 2.5. This change of parameters seems small but has a great impact on the time to carry out our simulations since solving the stochastic master equation with high precision requires many trajectories. In particular, when using the qutrit model (\ref{DHT3}) with the above new parameters, the dimension of the total Hilbert space becomes $33$ ($3$ states in qutrit and $11$ photons of cavity) instead of the dimension $16$ used in obtain Fig.~3 of the main text ($2$ states in qubit and $8$ photons in cavity). This means that each trajectory simulated with the stochastic master equation will take 120 seconds instead of 11 seconds, making the whole simulation of the QND detector tomography to last 10 times longer. Since doing our full QND detector tomography with the precision of Fig.~3 of the main text requires 20.000 individual trajectories for each of the 18 combinations of initial states and gates, our simulations took around 10 days, implying that the qutrit simulation will around 3 months. Optimizing such a simulation is unfeasible with the standard numerical methods used in this work, but it is possible to develop radically different and more efficient simulation techniques to provide a systematic study of QND readout of multi-level qubits in the near future. Independent of this, all these interesting non-dispersive effects can be immediately  experimentally explored by implementing our QND detector tomography procedure to a low anharmonic qubit such as a transmon.

\end{document}